\documentclass[a4paper,fleqn,usenatbib]{mnras}

\usepackage{newtxtext,newtxmath}

\usepackage[T1]{fontenc}
\usepackage{ae,aecompl}

\usepackage{graphicx}	
\usepackage{amsmath}	
\usepackage{amssymb}	

\title[CE mass-loss timescale constraint]{Constraints on the common-envelope evolution process from wide triple
systems}

\author[Erez Michaely and Hagai B. Perets]{
Erez Michaely,$^{1,2}$\thanks{E-mail: erezmichaely@gmail.com}
Hagai B. Perets$^{1}$
\\
$^{1}$Physics Department, Technion - Israel Institute of Technology,
Haifa 3200004, Israel\\
$^{2}$Astronomy Department, University of Maryland, College Park,
MD 20742\\
}

\date{Accepted XXX. Received YYY; in original form ZZZ}

\pubyear{2018}

\begin{document}
\label{firstpage}
\pagerange{\pageref{firstpage}--\pageref{lastpage}}
\maketitle

\begin{abstract}
Common envelope (CE) is an important phase in the evolution of interacting
evolved binary systems. The interaction of the binary components during
the CE evolution (CEE) stage gives rise to orbital inspiral and the
formation of a short-period binary or a merger, on the expense of
extending and/or ejecting the envelope. CEE is not well understood,
as hydrodynamical simulations show that only a fraction of the CE-mass
is ejected during the dynamical inspiral, in contrast with observations
of post-CE binaries. Different CE models suggest different timescales
are involved in the CE-ejection, and hence a measurement of the CE-ejection
timescale could provide direct constraints on the CEE-process. Here
we propose a novel method for constraining the mass-loss timescale
of the CE, using post-CE binaries which are part of wide-orbit triple
systems. The orbit/existence of a third companion constrains the CE
mass-loss timescale, since rapid CE mass-loss may disrupt the triple
system, while slower CE mass-loss may change the orbit of the third
companion without disrupting it. As first test-cases we examine two
observed post-CE binaries in wide triples, Wolf-1130 and GD-319. We
follow their evolution due to mass-loss using analytic and numerical
tools, and consider different mass-loss functions. We calculate a
wide grid of binary parameters and mass-loss timescales in order to
determine the most probable mass-loss timescale leading to the observed
properties of the systems. We  find that mass-loss timescales of the
order of $10^{3}-10^{5}{\rm yr}$ are the most likely to explain these
systems. Such long timescales are in tension with most of the CE mass-loss
models, which predict shorter, dynamical timescales, but are potentially
consistent with the longer timescales expected from the dust-driven
winds model for CE ejection.
\end{abstract}




\section{Introduction}

\label{sec:Introduction}

Common envelope evolution (CEE) is a relatively short binary evolution
phase, typically thought to occur on up to a few dynamical timescales.
It occurs when two stars orbit their center of mass in a shared CE
. CEE was first suggested by \citet{1976IAUS...73...75P} and \citet{1976IAUS...73...35V}
in order to explain the observations of short period (<days) binaries
with a degenerate component, now known as post-CE binaries. The degenerate
component in the binary must have been much greater in size than the
observed separation between the stars of the binary. Hence it was
suggested that both stars shared the large envelope of an evolved
star, and the dissipative interaction with the CE lead to the CE-ejection
and the inspiral of the binary into short-period, or even lead to
its merger. 

The driving engine of CE-ejection is still debated, and several competing
models were suggested (e.g. see \citealt{Ivanova2013} for a review,
see \citealt{Glanz2018} for a brief overview and updates). Hydrodynamical
simulations modeling the dynamical inspiral phase show that only a
fraction of the CE-mass is ejected during the inspiral \citep[e.g.][]{Taam2010,DeMarco2011,Passy2012,Ivanova2013}
raising a challenge to the simple picture of a dynamical inspiral,
and suggesting that additional source of energy are required for the
CE-ejection. The timescale for the fractional mass-loss is a few dynamical
timescales extending over a few up to a few tens of years. Ivanova
and collaborators (\citealt[and references therein]{Ivanova2015,Cla+17})
suggest that recombination energy from the cooling expanding envelope
can provide the necessary source, and their models suggest that the
CE-ejection time could be as long as of $10-100{\rm yr}$. \citet{Glanz2018}
suggest that radiation on dust forming in the outer layers of the
extended cooling envelope can provide a CE-mechanism similar to that
thought to operate in asymptotic giant branch (AGB) stars. In this
model the timescale for the CE-ejection is longer, likely $10^{4}-10^{5}{\rm yr}$,
and could be comparable to the timescale of AGB mass-loss. In principal,
measurement of the CE mass-loss timescale could constrain the different
models suggested. However, all of these various timescales are relatively
short in comparison with the stellar lifetimes, making the likelihood
for direct observations of the ejection small.

Here we propose a novel method to indirectly constrain the CE mass-loss
timescales using post-CE binaries which are part of wide triple systems.
The existence of a third wide-orbit companion to a post-CE binary,
and the orbit properties can potentially constrains the CE mass-loss
timescale, since rapid CE mass-loss may disrupt the triple system,
while slower CE mass-loss may change the orbit of the third companion
without disrupting it. Different mass-loss timescale and mass-loss
evolution give rise to different orbital configuration of the systems,
and the distribution of such wide triples can potentially constrain
CE-mass-loss timescales. 

Here we discuss the method and use two observed post-CE binaries in
triple systems as test-cases. We consider the two \textit{triple}
systems, Wolf 1130 \citep{Mace2018} and GD 319 \citep{Farihi2005};
both are post-CE binaries with distant third companions. Given the
wide separation (thousands of AU) we can treat outer orbit of the
triple system as a wide binary, and consider the inner post-CE binary
as a primary single point mass. The evolution and final outcome of
the outer-binary is determined by its initial configuration (masses,
separation and eccentricity) at the start of the mass loss \citep{Hills1983},
and the mass-loss evolution.

The paper is organized as follows: in section \ref{sec:Mass_loss_in_binary}
we discuss the mass-loss dynamics in binary systems, generally divided
to three timescale regimes: prompt (fast) mass loss, adiabatic (slow)
mass loss and comparable-timescale mass-loss. In section \ref{sec:Wolf1130}
we present the system Wolf 1130 and analyze its possible evolution
from the progenitor stage to its current configuration, as to find
the most probable mass loss timescale consistent with its currently
observed configuration. In section \ref{sec:GD319} we repeat the
analysis for the system GD 319 system. Finally, in section \ref{sec:Discussion}
we discuss the results and implications and then summarize (section
\ref{sec:Summary} ).

\section{Mass loss in binary systems}

\label{sec:Mass_loss_in_binary} 

The orbital elements (in particular the semi-major axis; SMA, $a$;
and eccentricity, $e$) of a gravitationally bound, but otherwise
non-interacting binary system are conserved. However, when the binary
losses mass the orbital elements may vary in time. We focus on three
different regimes, generally divided by ratio of the mass-loss times-scale,
$T_{{\rm ML}}$ to the orbital period of the binary, $P_{{\rm binary}}$.
These include the prompt mass-loss regime for which $T_{{\rm ML}}/P\ll1$
discussed in subsection \ref{subsec:prompt_massloss}; the adiabatic
mass-loss regime for which $T_{{\rm ML}}/P\gg1$ discussed in subsection
\ref{subsec:adiabatic_massloss}; and the regime of comparable timescales
$T_{{\rm ML}}\sim P$ discussed in subsection \ref{subsec:comparable}.
Throughout the discussion and analysis we assume an overall isotropic
mass loss and hence the \textit{specific} orbital angular momentum
is constant; future studies may explore more complex cases. Note the
total angular momentum in all of the cases changes due the mass loss.

\subsection{Prompt mass loss}

\label{subsec:prompt_massloss}

Prompt mass loss is defined by the following condition 
\begin{equation}
P_{{\rm binary}}\times\dot{M}_{{\rm binary}}\gg M_{{\rm binary}}\label{eq:prompt_definition}
\end{equation}
where $P_{{\rm binary}}$ is the binary orbital period, $M_{{\rm binary}}$
and $\dot{M}_{{\rm binary}}$ are the binary mass and the binary mass
loss respectively. Namely, in the regime of prompt mass loss a significant
fraction of the binary mass is ejected on timescales much shorter
than the orbital period. This case is relevant for fast events such
as supernova explosions or dynamical mass ejection. To first order,
in this regime we can approximate the mass loss to be instantaneous. 

Many studies explore prompt mass loss \citep[e.g. ][]{Blaauw1961,Hills1983}.
The ratio of the final SMA of the systems to its initial one in this
case is given by \citep{Hills1983}
\begin{equation}
\frac{a_{{\rm f}}}{a_{i}}=\frac{1}{2}\left[\frac{M_{{\rm binary}}-\Delta M}{\frac{1}{2}M_{{\rm binary}}-\left(\frac{a_{i}}{r}\right)\Delta M}\right]\label{eq:sma_ratio_promt}
\end{equation}
where $a_{i}$ and $a_{f}$ are the initial and final SMA respectively,
$\Delta M$ is the amount of mass lost from the binary and $r$ is
the separation of the components of the binary before the mass loss
occurred. An immediate result from (\ref{eq:sma_ratio_promt}) is
the condition for a binary to survive (i.e. still be bound) following
the mass-loss evolution in this regime is 
\begin{equation}
\frac{\Delta M}{M_{{\rm binary}}}\le\frac{r}{2a_{i}}.
\end{equation}

Therefore, a circular $\left(r=a_{i}\right)$ binary which loses more
than half of its original mass will be disrupted regardless of its
initial SMA. For an eccentric orbit the disruption of the binary depends
on the separation of the binary at the instantaneous moment of the
mass-loss event. The maximal separation for a binary with an eccentric
orbit is $r=a_{i}\left(1+e\right)$, and the condition for binary
survival is 
\begin{equation}
\alpha\le\frac{1+e}{2}
\end{equation}
where $\alpha$ is the fraction of mass lost from the system. Furthermore,
the final eccentricity is given by 
\begin{equation}
e_{f}=\left\{ 1-\left(1-e_{i}^{2}\right)\left[\frac{1-\left(\frac{2a_{i}}{r}\right)\left(\frac{\Delta M}{M_{{\rm binary}}}\right)}{\left(1-\frac{\Delta M}{M_{{\rm binary}}}\right)^{2}}\right]\right\} ^{1/2}.
\end{equation}

\subsection{Adiabatic mass loss}

\label{subsec:adiabatic_massloss}

Adiabatic mass loss is defined by the following condition:
\begin{equation}
P_{{\rm binary}}\times\dot{M}_{{\rm binary}}\ll M_{{\rm binary}}.\label{eq:adiabatic_definition}
\end{equation}
Namely, the amount of mass lost from the system in one orbit is negligible
compared to the total mass of the system. In this case, where the
mass-loss timescale is much longer than the orbital period we can
treat the system with its adiabatic invariant and get the following
well known results for the SMA and eccentricity \citep{Jeans1961}:
\begin{equation}
\frac{a_{f}}{a_{i}}=\frac{M_{{\rm binary}}}{M_{{\rm binary}}-\Delta M}\label{eq:sma_ratio_adiabatic}
\end{equation}
 
\begin{equation}
e_{f}=e_{i}.
\end{equation}
In this case the system remains bound irrespectively of the total
mass lost from the system period and only expands its separation while
keeping a constant eccentricity.

\subsection{Comparable timescales mass loss and period}

\label{subsec:comparable}

In the previous two subsections we described the orbital evolution
of two extreme cases where the mass-loss timescale is much longer
(adiabatic) or much shorter (prompt) than the orbital period. The
evolution in both of these regimes can be well described by an approximate
analytic expression for change in the orbital elements. In this subsection
we briefly discuss the third regime in which the mass-loss timescale
is comparable to the orbital period, namely:
\begin{equation}
P_{{\rm binary}}\times\dot{M}_{{\rm binary}}\approx M_{{\rm binary}}.\label{eq:comprable_definition}
\end{equation}

In order to determine the evolution of the orbital elements in this
case we need to resort to numerical integration. We note that unlike
in the prompt and adiabatic regime the evolution depends on the prescribed
detailed mass-loss evolution function, and not only on the overall
change in mass and the initial position of the binary. Hence one needs
to prescribe a specific mass-loss function
\begin{equation}
\dot{M}_{{\rm binary}}=M\left(t\right).
\end{equation}
Since the exact function of the mass-loss in CE-ejection in this mass-loss
range is not known specifically, we consider two general mass-loss
functions, either a linear mass-loss rate or an exponential one, as
we describe in the next section.

\section{The case of a Wolf 1130}

\label{sec:Wolf1130}

Wolf 1130 \citep{Mace2013,Mace2018} is a nearby triple systems (distance
of $16.7\text{\textpm}0.2{\rm pc}$). The inner binary consists of
an M subdwarf (Wolf 1130A) with mass $\sim0.3M_{\odot}$ and a white
dwarf (Wolf 1130B) with mass of $\sim1.242M_{\odot}$ \citep{Mace2018}.
The orbital period of Wolf 1130AB is $P=0.4967{\rm day}$. The tertiary
(Wolf 1130C) is a subdwarf brown dwarf with mass $\sim0.05M_{\odot}.$
Wolf 1130C has a common proper motion to Wolf 1130AB. The projected
separation between Wolf 1130B and Wolf 1130AB is of $\sim3150{\rm AU}$
\citep{Mace2013}.

Wolf 1130AB is a post-CE binary, we used \texttt{the binary\_C} code
\citep{Izzard2004,Izzard2006,Izzard2009} to find a zero age main
sequence binary that will evolve to form the observed Wolf 1130AB.

We found the following system: both starts with Solar metallicity
of $0.016$, the primary with $m_{1}=7.2M_{\odot}$ and the secondary
with $m_{2}=0.3M_{\odot}.$ The initial circular orbital separation
is $a=2000R_{\odot}\approx9.3{\rm AU}.$ We used the parameters of
\texttt{binary\_C} for the relevant parameters presented in Table
\ref{tab:binary_C-parameters-used}. After $52.5{\rm Myr}$ CEE begins
when the primary loses $\sim1M_{{\rm \odot}}$ to become a $\sim6.3M_{{\rm \odot}}.$
At the end of the CE phase the white dwarf has a mass of $\sim1.3M_{\odot}$
and the binary period is $P\approx0.5{\rm days}$, similar to the
infer parameters of the observed Wolf 1130AB system.

\subsection{Numerical calculation}

\label{subsec:Numerical_calc_Wolf}

In order to constrain the mass loss timescale we use an N-body integrator
with a variable time step, using the Hermite fourth order integration
scheme following \citep{Hut1995}. We adjusted the code to include
mass loss from the primary. We initialized the binary as follows.
The primary (the post-CE binary treated now as a point mass) with
initial mass of $m_{1i}=6.6M_{\odot}$, and final mass (after the
CE mass-loss) of $m_{1f}=1.6M_{{\rm \odot}}.$ The secondary with
$m_{2}=0.3M_{\odot}$. We run a grid of simulations with the following
parameters: the initial eccentricity $e_{i}$, sampling over $9$
values from $0.1$ to $0.9$ equally spaced, the initial SMA, where
we considered the following separations $a_{i}=30,60,100,150,200,300{\rm AU}$;
and a range of equally log-spaced mass-loss timescales, $t_{{\rm ML}}=1,10,10^{2},10^{3},10^{4},10^{5}{\rm yr}$.
For each of the points on our grid we consider $1000$ equally spaced
values for the mean anomaly , ${\scriptstyle M}$, in the range $0-2\pi$. 

For each simulation we considered two possible mass-loss functional
forms. The first mass-loss scheme is linear with time 
\begin{equation}
\dot{M}_{{\rm binary}}=-\beta\label{eq:linear mass loss}
\end{equation}
where $\beta$ is set by $\beta=\left(m_{1f}-m_{1i}\right)/t_{{\rm ML}}$
for each mass loss timescale. 

The second mass-loss scheme is exponential 
\begin{equation}
M\left(t\right)_{{\rm binary}}=m_{1i}e^{-t\gamma}\label{eq:exponential}
\end{equation}
where $\gamma=-\ln\left(m_{1f}/m_{1i}\right)/t_{{\rm ML}}$. 

Additionally, in order to save computer time we did not simulate the
combinations of SMAs and eccentricities that are unstable on a dynamical
time scale. For each simulation we calculate the final average separation
\begin{equation}
s_{f}\equiv a_{f}\left(1+\frac{1}{2}e_{f}^{2}\right).
\end{equation}
\citet{Dupuy2011} calculated the statistical conversion factor from
a binary orbit SMA, $a$, to the observed projected separation, $\rho$,
for different eccentricity distributions. Following their results
we use the lower and upper conversion factors for one standard deviation,
$1\sigma$. We consider two possible cases. In case of a uniform distribution
of the eccentricities the conversion factors are 0.75 and 2.02 for
the lower and upper bounds, respectively, which translate to $2362.5{\rm AU}$
and $6363{\rm AU}$. For the thermal eccentricity distribution the
conversion factors are $\left(0.67,2.07\right)$, which translate
to $\left(2110.5{\rm AU,6520.5{\rm AU}}\right)$. Hence, in order
for the final orbit to be consistent with the observed system it needs
to reside in the range
\[
2362.5{\rm AU}<s_{f}<6363{\rm AU},
\]
assuming a uniform initial eccentricity distribution, or in the range
\begin{equation}
2110.5{\rm AU}<s_{f}<6520.5{\rm AU},\label{eq:condition_thermal_wolf}
\end{equation}
 assuming a thermal distribution of the initial eccentricity.

We follow the evolution of the each of systems sampled in our grid,
to find all the cases which end-up up in the appropriate range, and
record the number of successful cases among the initial $1000$ different
mean anomalies. 

In order to asses the likelihood for a CE-ejection timescale to provide
consistent systems, we need to consider some a prior distributions
for the initial system parameters as to weigh them correctly. We consider
two possible initial eccentricity distributions for this purpose,
either uniform in $e$ or a thermal distribution namely, $f\left(e\right)\propto e$
\citep{Tokovinin2016,Moe2017}. For both distributions we normalized
the function with $e_{{\rm min}}=0$ and $e_{{\rm max}}=0.9$. Next
we need to convolute with the SMA initial distribution, for which
we assume the well known Opik law, specifically $f\left(a\right)\propto1/a$
which is uniform in log space, normalized by $a_{{\rm min}}=20{\rm AU}$
and $a_{{\rm max}}=300{\rm AU}.$ Equipped with these plausible assumptions
we can now evaluate the relative likelihood for the CE mass-loss timescale.
In subsection \ref{subsec:linear} we present the result for linear
mass loss function (for both eccentricity distributions) while in
subsection \ref{subsec:exponential} we present the results for the
exponential mass loss.

\begin{table}
\caption{\label{tab:binary_C-parameters-used}\texttt{binary\_C} parameters
used to evolve the simulated zero age main sequence binary into the
observed Wolf 1130AB.}

\begin{tabular}{c|c|c|c|c|c|c}
\hline 
\multicolumn{7}{c}{binary\_C parameters for Wolf 1130AB}\tabularnewline
\hline 
\hline 
$m_{1}\left[M_{\odot}\right]$ & $m_{2}\left[M_{\odot}\right]$ & $a\left[R_{\odot}\right]$ & $e$ & $\alpha_{{\rm CE}}$ & $\lambda_{{\rm CE}}$ & $\lambda_{{\rm ion}}$\tabularnewline
\hline 
\hline 
$7.2$ & $0.3$ & $2000$ & $0$ & $0.5$ & $-1$ & $0$\tabularnewline
\hline 
\end{tabular}
\end{table}

\subsubsection{The case of a linear mass-loss evolution}

\begin{figure*}
\includegraphics[width=1\columnwidth]{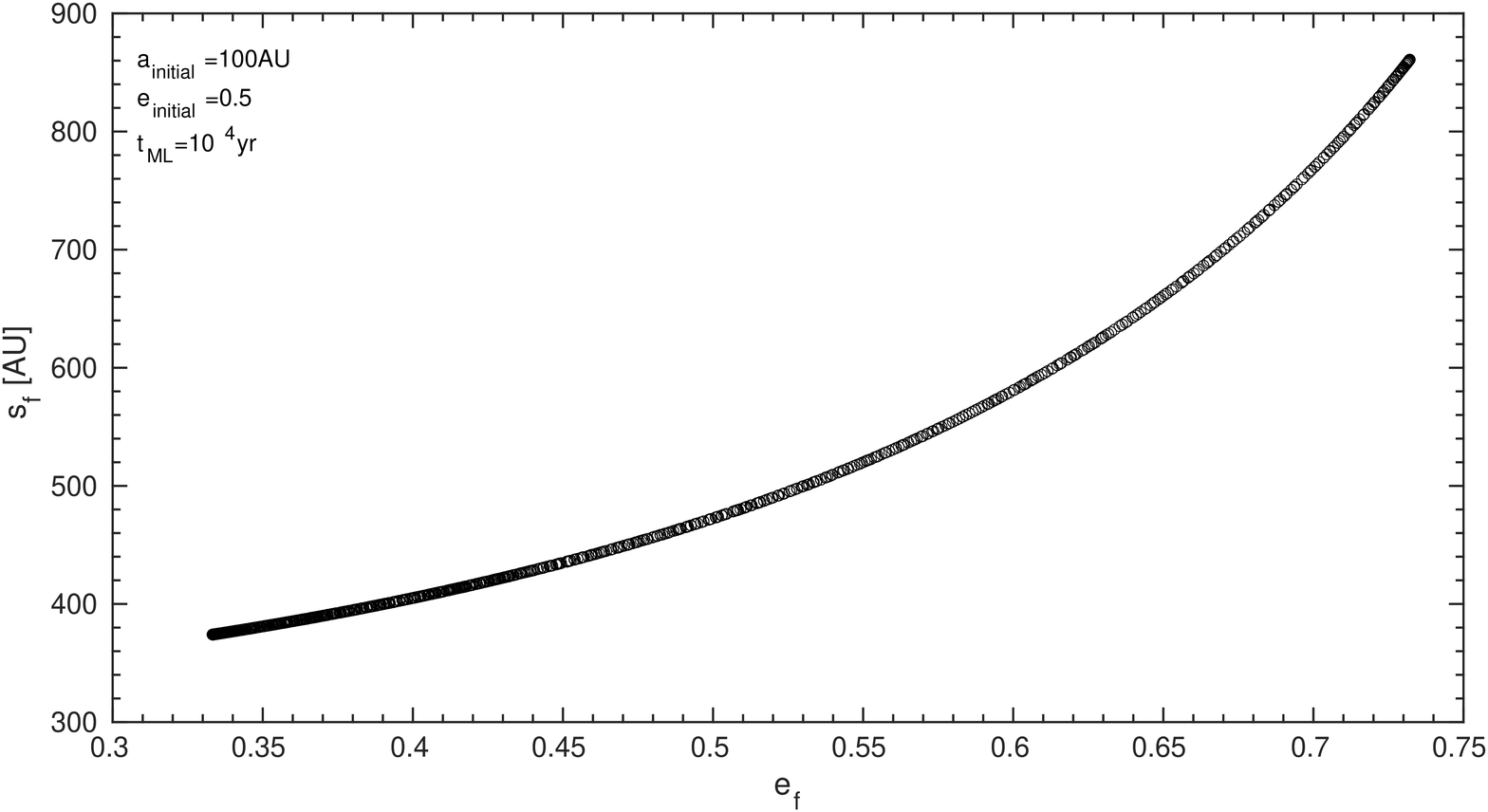}\includegraphics[width=1\columnwidth]{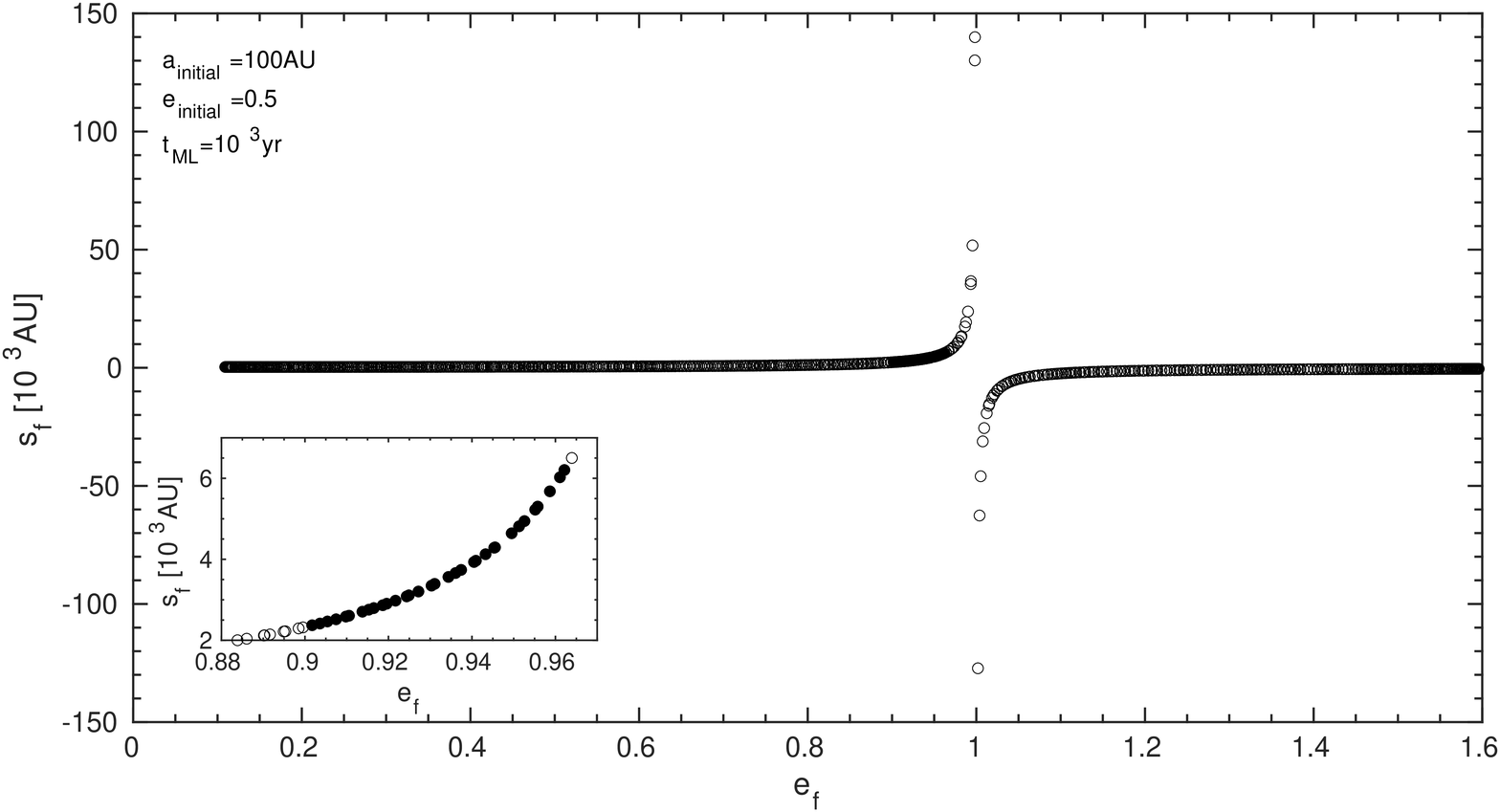}\caption{\label{fig:example_linear}Example of the simulated data. Left plot:
The simulated system is $m_{1i}=6.6M_{\odot}$, $m_{{\rm 1f}}=1.6M_{\odot}$
, $m_{2}=0.3M_{\odot}$, $a_{i}=100{\rm AU}$, $e_{i}=0.5$ and $T_{{\rm ML}}=10^{4}{\rm yr}$
with a linear mass loss function. The average final separation, $s_{f}$
is plotted against the final eccentricity for $1000$ equally spaced
initial mean anomaly. The maximal $s_{f}$ is below $900{\rm AU}$
hence not satisfying condition (\ref{eq:condition_thermal_wolf}).
Right plot: the same systems as left plot only with $T_{{\rm ML}}=10^{3}{\rm yr}.$
In this case fraction of the initial values of the mean anomaly result
in a binary breakup, $e_{f}>1$ and $s_{f}<0$. The zoomed in caption
is the small fraction of the initial mean anomaly that satisfies the
conditions (\ref{eq:condition_thermal_wolf}).}

\end{figure*}

For the linear mass loss case we calculate the fraction of systems
that satisfy conditions (\ref{eq:condition_thermal_wolf}) for each
initial SMA, $a_{i}$, initial eccentricity, $e_{i}$ and mass loss
timescale $T_{{\rm ML}}$ out of $1000$ equally space mean anomalies,
${\scriptscriptstyle M}\in\left\{ 0,2\pi\right\} $. As an example
we present the results of two such simulations in Figure \ref{fig:example_linear}.
For $a_{i}=100{\rm AU}$, $e_{i}=0.5$ and mass loss timescale of
$T_{{\rm ML}}=10^{4}{\rm yr}$ the upper plot of Figure \ref{fig:example_linear}
presents the average final separation, $s_{f}$ as a function of final
eccentricity, $e_{f}$. In this example none of the case ends up satisfying
conditions (\ref{eq:condition_thermal_wolf}). The lower plot of Figure
\ref{fig:example_linear} is similar to the upper plot but correspond
to the $T_{{\rm ML}}=1000{\rm yr}$ case. In this case a significant
fraction of tuns ends up in a disrupted binary, namely a negative
SMA or $e_{f}>1$. The remaining data points indicate a surviving
binary and a fraction of those satisfy conditions (\ref{eq:condition_thermal_wolf}).
The subplot figure presents the fraction that satisfy conditions (\ref{eq:condition_thermal_wolf})
depicted by full black markers. 

For each $a_{i}$ we compute a function, $F_{a_{i}}\left(e_{i},T_{{\rm ML}}\right)$,
which is the fraction of systems that satisfy conditions (\ref{eq:condition_thermal_wolf})
as a function of $e_{i}$ and $T_{{\rm ML}}$. Table \ref{tab:frac_example_100AU_linear}
show one example of $F_{a_{i}}\left(e_{i},T_{{\rm ML}}\right)$ for
$a_{i}=100{\rm AU}$, for a uniform distribution of the initial eccentricities
and assuming a linear mass-loss. 

\begin{figure}
\includegraphics[width=1\columnwidth]{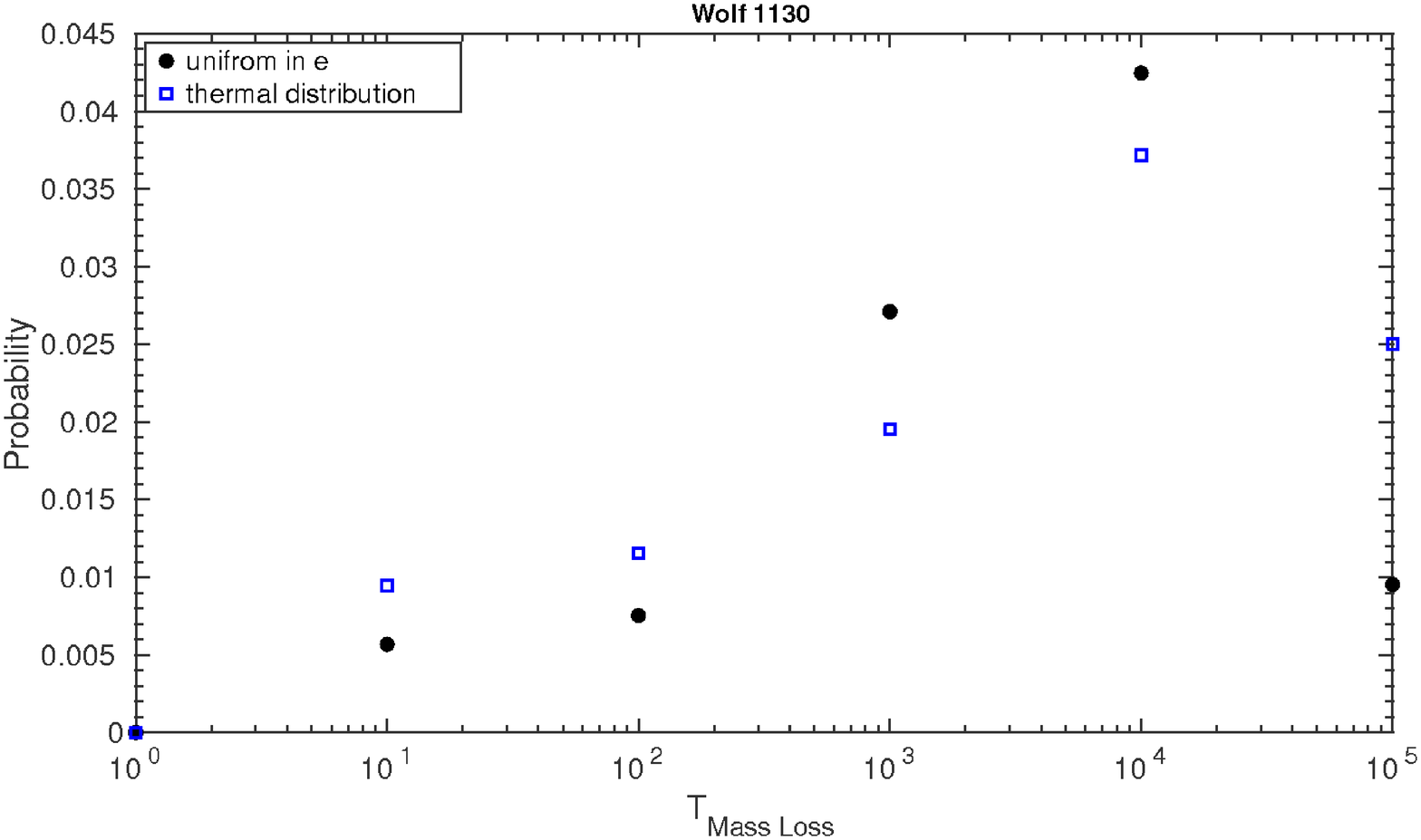}

\includegraphics[width=1\columnwidth]{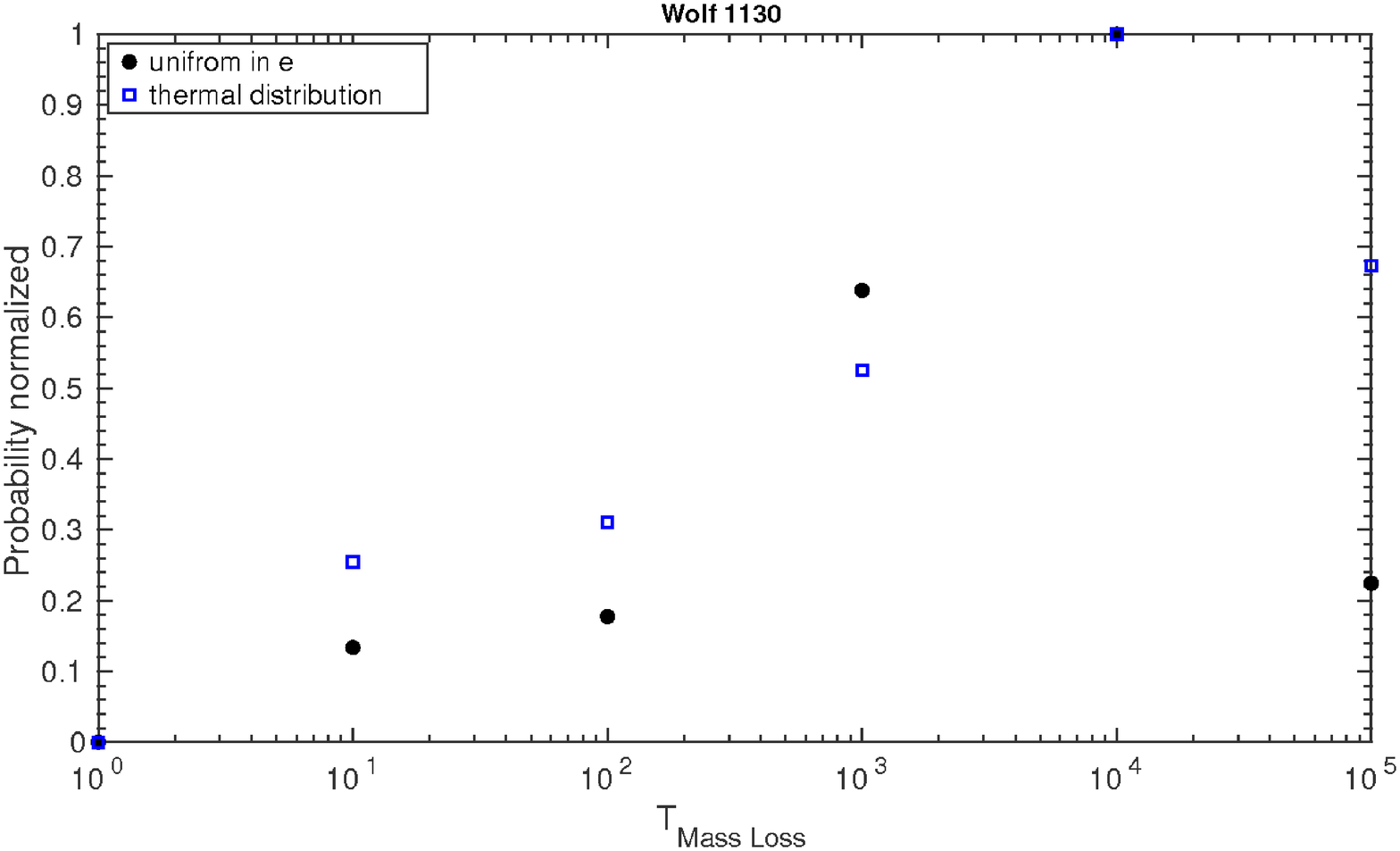}\caption{\label{fig:Probability_linear}Upper plot: the probability of the
mass loss timescale from the observed Wolf 1130 system for the linear
mass loss case. The black circles (blue squares) correspond to uniform
distribution (thermal distribution) of initial eccentricity. The most
probable timescale is $T_{{\rm ML}}=10^{4}{\rm yr}$. Bottom plot:
Is the normalized version of the upper plot to the maximal value for
each distribution, in both cases is $T_{{\rm ML}}=10^{4}{\rm yr}$.}
\end{figure}

Figure (\ref{fig:Probability_linear}) presents the mass-loss timescale
constraints for Wolf 1130. Black circles and blue squares represent
uniform distribution and thermal distribution of the initial eccentricities,
respectively. For both distributions the most probable timescale is
$T_{{\rm ML}}=10^{4}\text{{\rm yr}}.$ For the uniform distribution
case we find that $T_{{\rm ML}}=10^{4}{\rm yr}$ is more probable
than $10^{5}{\rm yr,}10^{3}{\rm yr},10^{2}{\rm yr},10{\rm yr}$ by
a factor of $4.08,1.57,5.64,7.48$, respectively. For the thermal
distribution case we find that $T_{{\rm ML}}=10^{4}{\rm yr}$ is more
probable than $10^{5}{\rm yr},10^{3}{\rm yr},10^{2}{\rm yr},10{\rm yr}$
by a factor of approximately $1.49,1.9,3.2,4$, respectively as shown
in the bottom plot of Figure (\ref{fig:Probability_linear}). For
both eccentricity distribution we find zero probability for $T_{{\rm ML}}=1{\rm yr}.$

\label{subsec:linear}

\begin{table}
\caption{\label{tab:frac_example_100AU_linear}Example of the simulated function
$F_{a_{i}}\left(e_{i},T_{{\rm ML}}\right)$ for $a_{i}=100{\rm AU}$
and for linear mass loss function and uniform initial eccentricity
distribution. In this case only $T_{{\rm ML}}=10^{3}{\rm yr}$ satisfied
the condition (\ref{eq:condition_thermal_wolf}). For this set of
initial conditions we did not simulate $e_{i}>0.5$ because for these
systems the pericenter is too close to the inner binary orbit and
the triple stability criteria breaks.}

\begin{tabular}{c|c|c|c|c|c}
\hline 
\multicolumn{6}{c}{Example for the value of $F_{a_{i}}\left(e_{i},T_{{\rm ML}}\right)$
for $a_{i}=100{\rm AU}$ }\tabularnewline
\hline 
fractions & $e=0.1$ & $e=0.2$ & $e=0.3$ & $e=0.4$ & $e=0.5$\tabularnewline
\hline 
\hline 
$T_{{\rm ML}}=10{\rm yr}$ & $0$ & $0$ & $0$ & $0$ & $0$\tabularnewline
\hline 
$T_{{\rm ML}}=100{\rm yr}$ & $0$ & $0$ & $0$ & $0$ & $0$\tabularnewline
\hline 
$T_{{\rm ML}}=1000{\rm yr}$ & $0.308$ & $0.09$ & $0.056$ & $0.042$ & $0.033$\tabularnewline
\hline 
$T_{{\rm ML}}=10000{\rm yr}$ & $0$ & $0$ & $0$ & $0$ & $0$\tabularnewline
\hline 
\end{tabular}

\end{table}

\subsubsection{The case of an exponential mass-loss evolution}

\label{subsec:exponential}

\begin{figure}
\includegraphics[width=1\columnwidth]{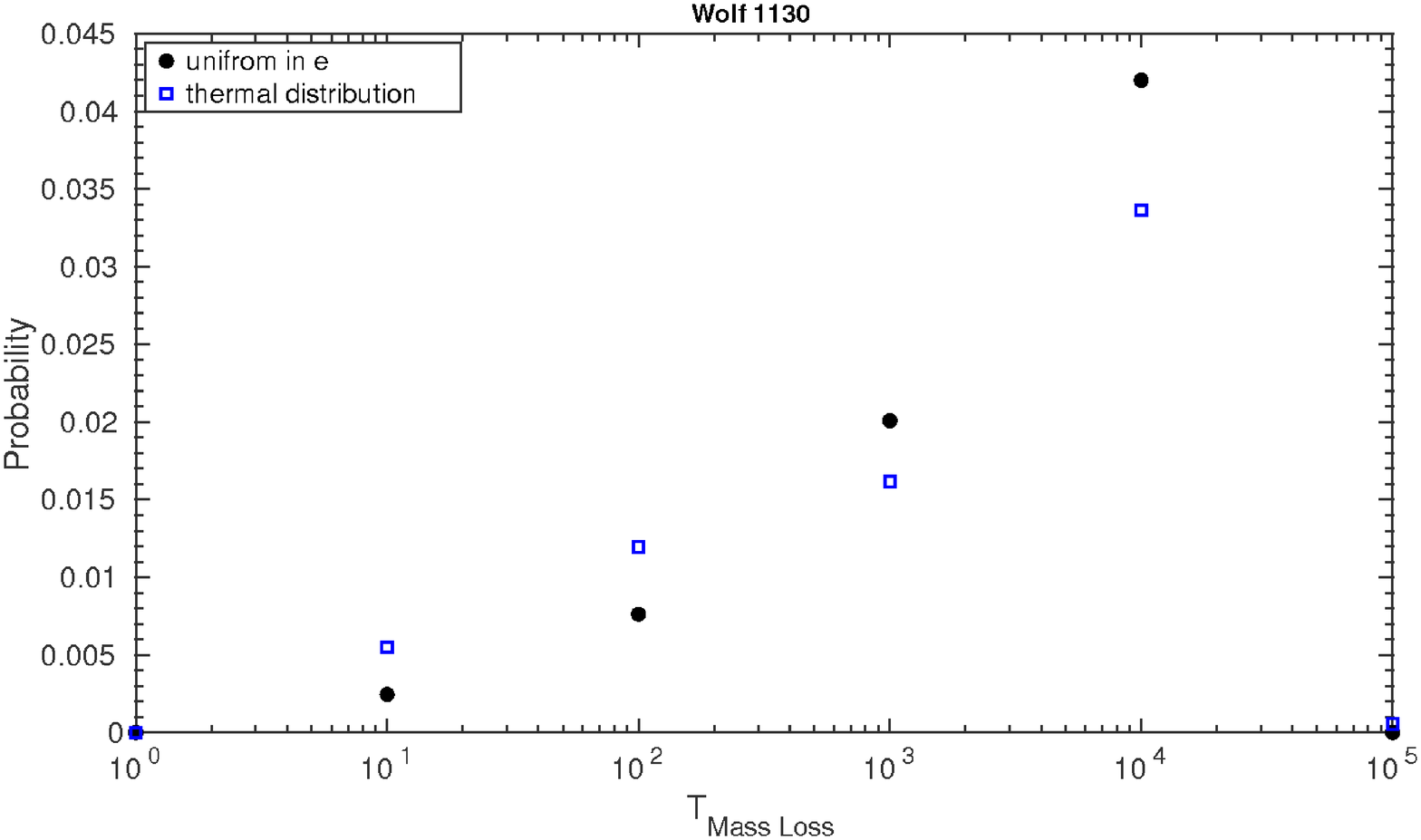}

\includegraphics[width=1\columnwidth]{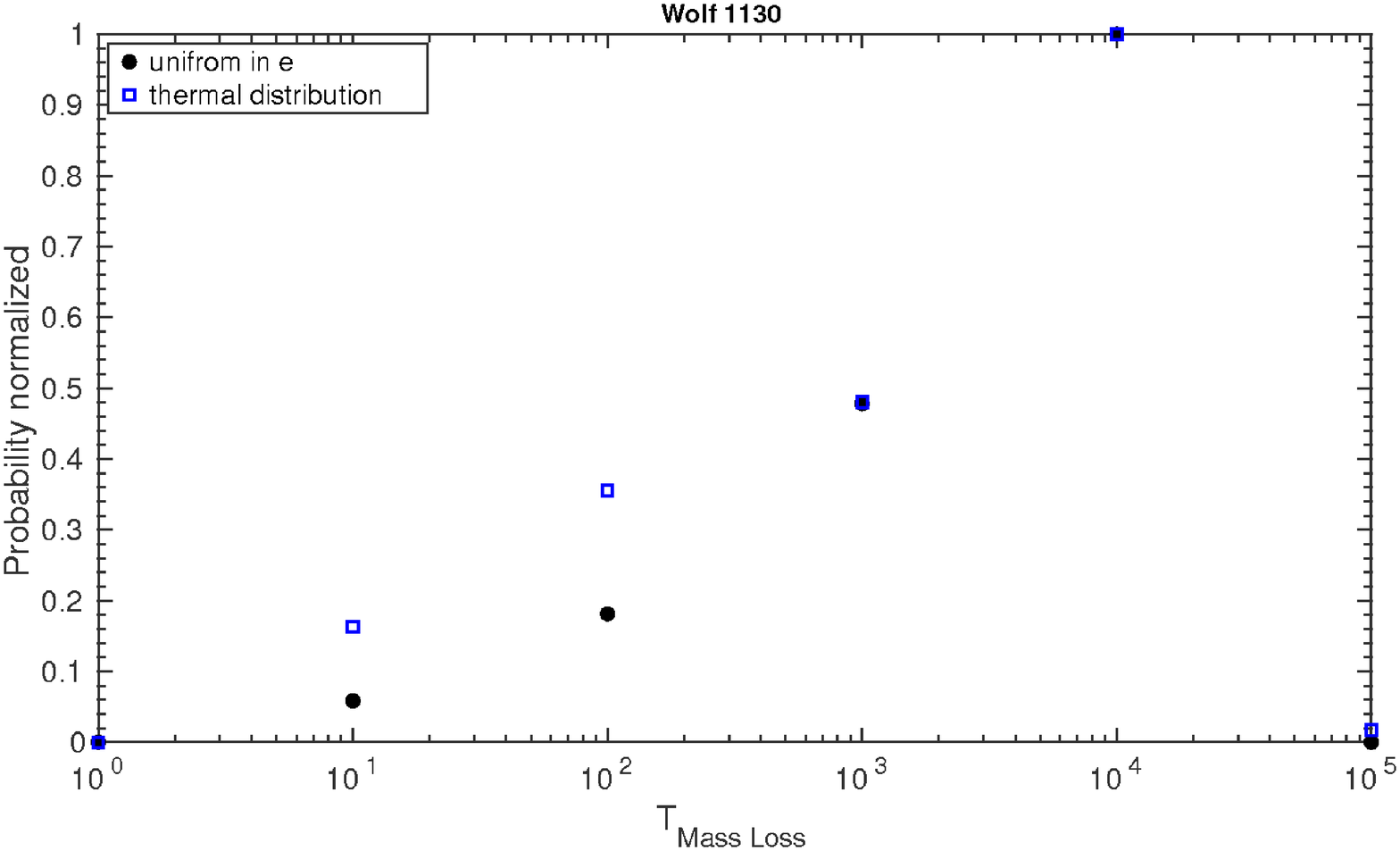}

\caption{\label{fig:wolf_exponential}Same as Figure \ref{fig:Probability_linear}.
Also for the exponential mass loss case the most probable mass loss
timescale is $T_{{\rm ML}}=10^{4}{\rm yr}$ for the system Wolf 1130.}

\end{figure}

In this subsection we present the results for the exponential mass
loss scenario, namely $M\left(t\right)_{{\rm binary}}=m_{1i}e^{-t\gamma}$
where $\gamma=-\ln\left(m_{1f}/m_{1i}\right)/t_{{\rm ML}}$. We follow
the same procedure as described in subsection \ref{subsec:linear}.
Figure \ref{fig:wolf_exponential} presents the probability (upper)
and the normalized probability to the maximal value (bottom). The
same as the linear mass loss scenario the most probable mass loss
timescale is $T_{{\rm ML}}=10^{4}{\rm yr}.$ For the uniform eccentricity
distribution this timescale is more probable than $10^{3}{\rm yr},10^{2}{\rm yr},10{\rm yr}$
by a factor of $2.1,5.6,17.2$ respectively. Moreover, we find zero
probability for $T_{{\rm ML}}=1{\rm yr}$ and $T_{{\rm ML}}=1{\rm 0^{5}yr}$.
While for the thermal distribution case this timescale is more probable
than $10^{5}{\rm yr,}10^{3}{\rm yr},10^{2}{\rm yr,10{\rm yr}}$ by
a factor of approximately $60,2.1,2.8,6.3$ respectively. For the
thermal eccentricity distribution we find zero probability for $T_{{\rm ML}}=1{\rm yr}.$

\section{Case of GD 319}

\label{sec:GD319}

GD 319 is a triple system \citep{Saffer1998,Farihi2005} at a distance
of $\sim436.5{\rm pc}.$ The inner binary consists of a subdwarf B
star (sdB) of mass $\sim0.5M_{\odot}$ with a close unseen companion
\citep{Saffer1998} of type dM. The tertiary is an M3.5 dwarf with
mass $\sim0.25M_{\odot}$ \citep{Farihi2005} at a separation of $\sim54914{\rm AU}$
from the inner binary with a common proper motion. The inner orbit
has a $0.6$ day period. 

As in the previous case \texttt{binary\_C} code by was used to find
a zero age main sequence binary that will evolve to form the observed
GD 319. We find the following best fir parameters: for a Solar metallicity
the primary begins with $m_{1}=1.5M_{\odot}$ and the secondary with
$m_{2}=0.5M_{\odot}.$ The initial circular orbital separation is
$a=350R_{\odot}$, as listed in Table \ref{tab:binary_C-parameters-used-GD319}.
After $2.8{\rm Gyr}$ CEE commences. At the end of CE phase the sdB
has a mass of $\sim0.5M_{\odot}$ and the binary period is $P\approx0.5$
day, consistent with the inferred parameters the observed GD 319 system. 

\begin{table}
\caption{\label{tab:binary_C-parameters-used-GD319}\texttt{binary\_C} parameters
used to evolve the simulated zero age main sequence binary into the
observed GD 319AB.}

\begin{tabular}{c|c|c|c|c|c|c}
\hline 
\multicolumn{7}{c}{\texttt{binary\_C} parameters for GD 319AB}\tabularnewline
\hline 
\hline 
$m_{1}\left[M_{\odot}\right]$ & $m_{2}\left[M_{\odot}\right]$ & $a\left[R_{\odot}\right]$ & $e$ & $\alpha_{{\rm CE}}$ & $\lambda_{{\rm CE}}$ & $\lambda_{{\rm ion}}$\tabularnewline
\hline 
\hline 
$1.5$ & $0.5$ & $350$ & $0$ & $0.2$ & $-1$ & $0$\tabularnewline
\hline 
\end{tabular}
\end{table}

\subsection{Numerical calculation}

We initiated the primary (the post-CE binary now treated as a point
mass) with initial mass of $m_{1i}=2M_{\odot}$ and final mass following
CEE of $m_{1f}=1M_{{\rm \odot}}.$ The secondary mass is $m_{2}=0.25M_{\odot}$.
We repeat the same type of analysis as described for the Wolf 1130
system, using the same sampling of parameters and mass-loss functions.
We only changes the separation and mass-loss timescale ranges to $a_{i}=50,100,300,500,1000,10000{\rm AU}$
and $t_{{\rm ML}}=1,10,10^{2},10^{3},10^{4},10^{5},10^{6}{\rm yr},$
respectively, corresponding to the wider separation of this system
and the longer period of the outer orbit. 

For this system we search for the following conditions \citep{Dupuy2011}.
For a uniform distribution of the initial eccentricities
\[
4.118\cdot10^{4}{\rm AU}<s_{f}<11.1\cdot10^{4}{\rm AU}
\]
and for the thermal distribution we use 
\begin{equation}
3.78\cdot10^{4}{\rm AU}<s_{f}<11.3\cdot10^{4}{\rm AU}.\label{eq:condition-GD319}
\end{equation}
 The rest of the numerical treatment is identical to the one described
at subsection \ref{subsec:Numerical_calc_Wolf}.

\subsubsection{The case of a linear mass-loss evolution}

\label{subsec:LinearMassLossGD319}

\begin{figure}
\includegraphics[width=1\columnwidth]{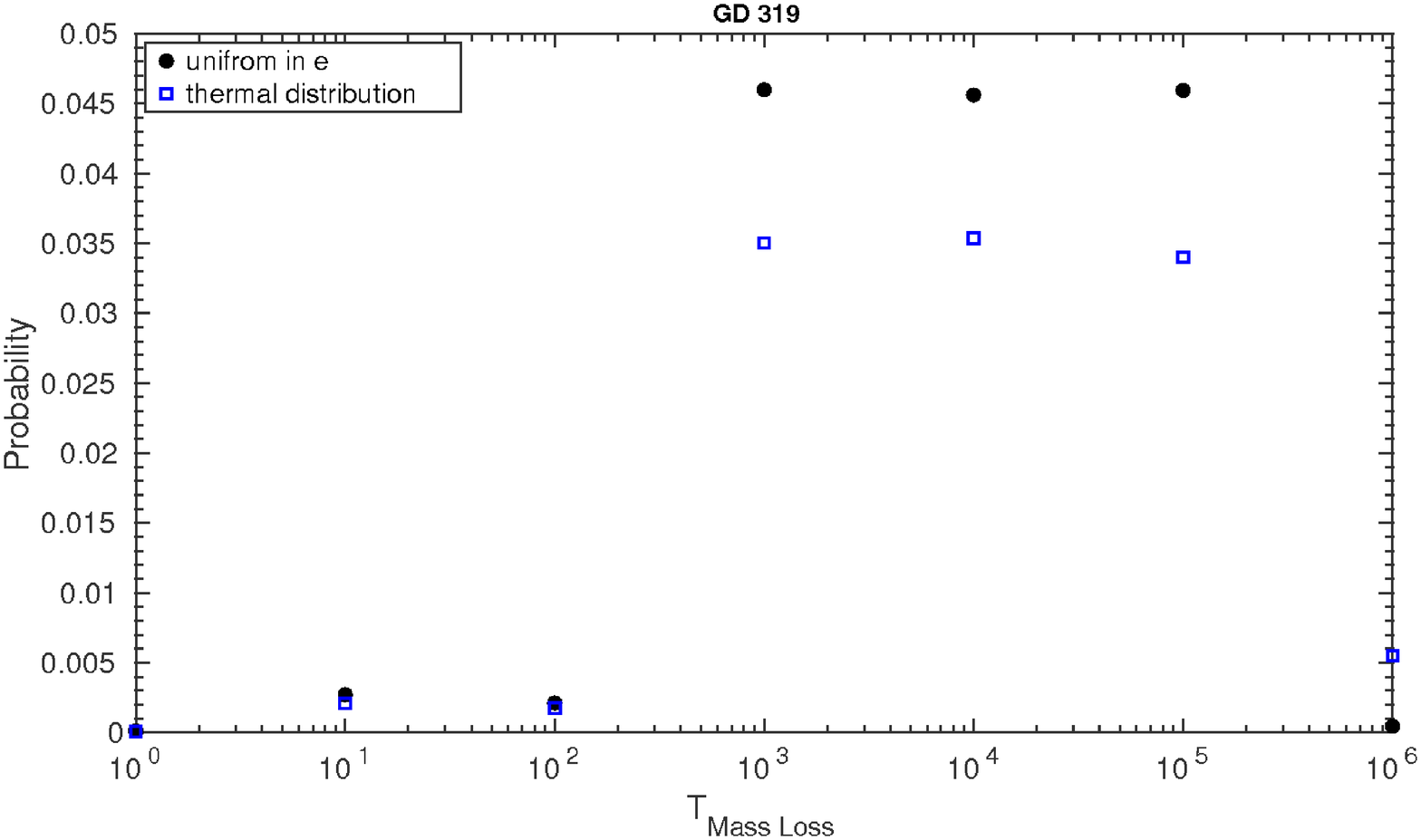}

\includegraphics[width=1\columnwidth]{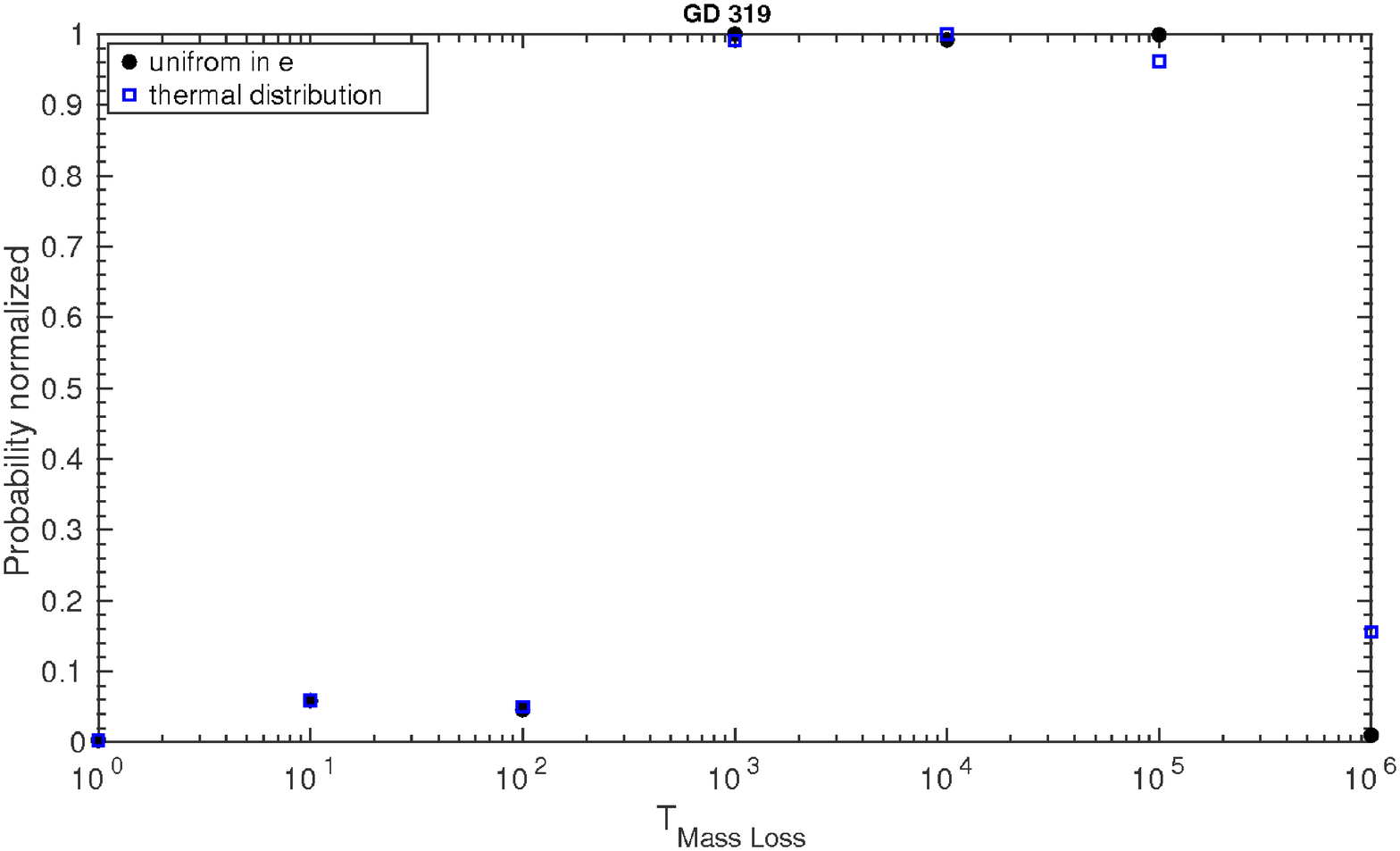}\caption{\label{fig:Linear-mass-loss_GD319}Upper plot: the probability of
the mass loss timescale from the observed GD 319 system for the linear
mass loss case. The black circles (blue squares) correspond to uniform
distribution (thermal distribution) of initial eccentricity. The most
probable timescale for the uniform initial eccentricity is $T_{{\rm ML}}=10^{3}{\rm yr}$
is $T_{{\rm ML}}=10^{4}{\rm yr}$ for while for the thermal eccentricity
distribution is $T_{{\rm ML}}=10^{4}{\rm yr}$. Bottom plot: Is the
normalized version of the upper plot to the maximal value for each
distribution. It is clear that mass loss timescales of $10^{3}-10^{5}{\rm yr}$
are favorable compared to $1-10^{2}{\rm yr}$ and $10^{6}{\rm yr}.$ }
 
\end{figure}

In this subsection we present the likelihood results of the mass loss
timescales for the linear mass loss function. In Figure \ref{fig:Linear-mass-loss_GD319}
(upper) we present the likelihood for each timescale, black circles
and blue squares represent uniform distribution and thermal distribution
of the initial eccentricity, respectively. The bottom panel of Figure
\ref{fig:Linear-mass-loss_GD319} shows the relative likelihood, namely
the ratio of the probabilities of all the calculated timescales to
the most probable one. 

We find that for the uniform eccentricity distribution the most probable
timescale is $T_{{\rm ML}}=10^{3}$, which is as likely as the case
of $T_{{\rm ML}}=10^{4}{\rm yr}$ and $T_{{\rm ML}}=10^{5}{\rm yr}$
. This timescale is more probable by a factor of approximately $17$,
$21.7$ and 100 than the case of $10{\rm yr}$, $100{\rm yr}$ and
$10^{6}$, respectively. Unlike the similar likelihood for the $10^{4}{\rm yr}$
and $10^{5}{\rm yr}$ timescales. 

For the thermal distribution of the initial eccentricity we find similar
results. The most probable mass loss timescale is $T_{{\rm ML}}=10^{4}$.
This timescale is significantly more likely than the $10{\rm yr}$,
$100{\rm yr}$ or $10^{6}$ yrs cases by factors of approximately
$17,20,6.7$, respectively. While the mass loss timescale of $10^{3}{\rm yr}$
and $10^{5}{\rm yr}$ share the likelihood of a factor of $1.001$
and $1.04$ respectively. The probability of $T_{{\rm ML}}=1{\rm yr}$
is zero for both initial eccentricity distributions.

\subsubsection{The case of an exponential mass-loss evolution}

\label{subsec:ExpMassLossGD319}

\begin{figure}
\includegraphics[width=1\columnwidth]{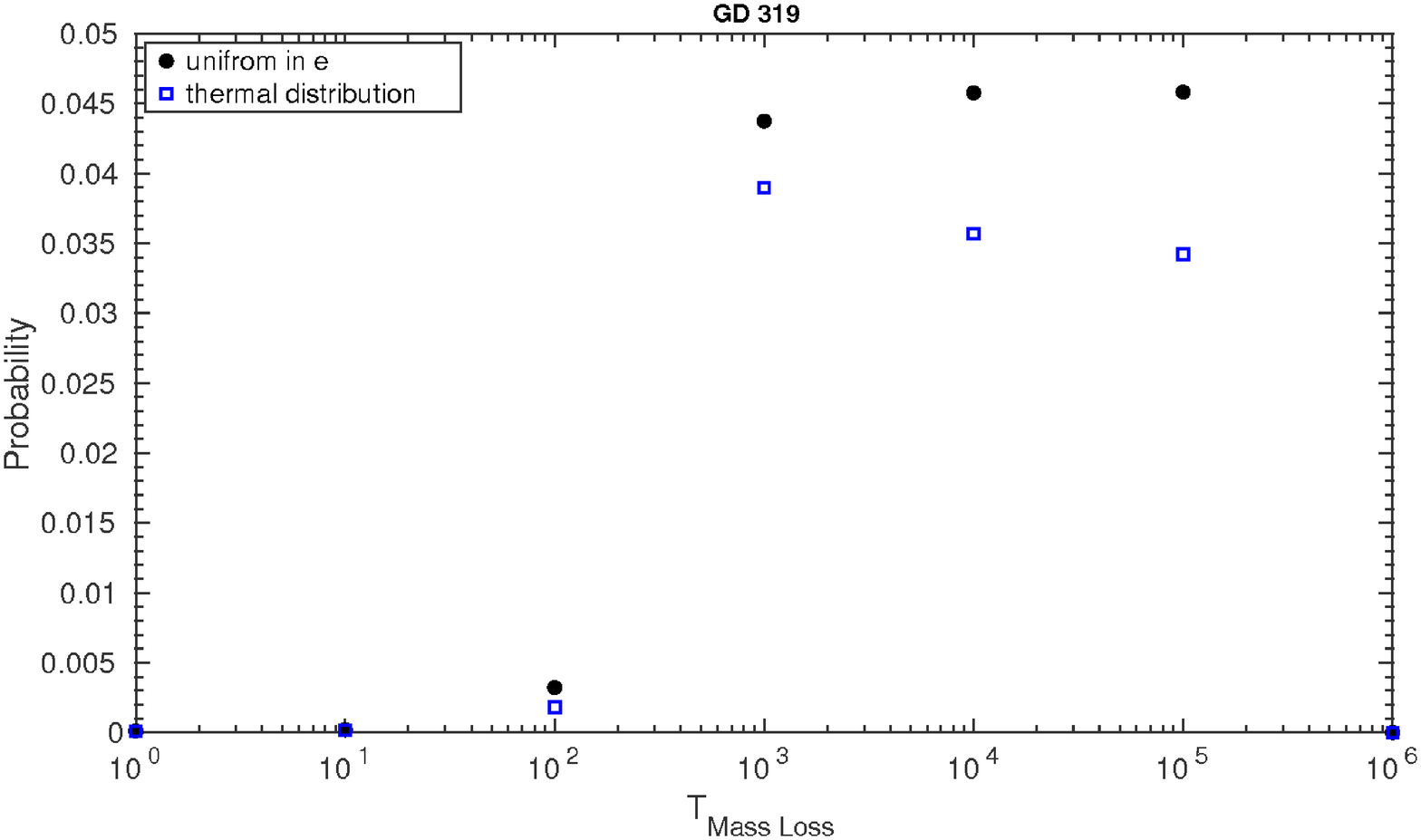}

\includegraphics[width=1\columnwidth]{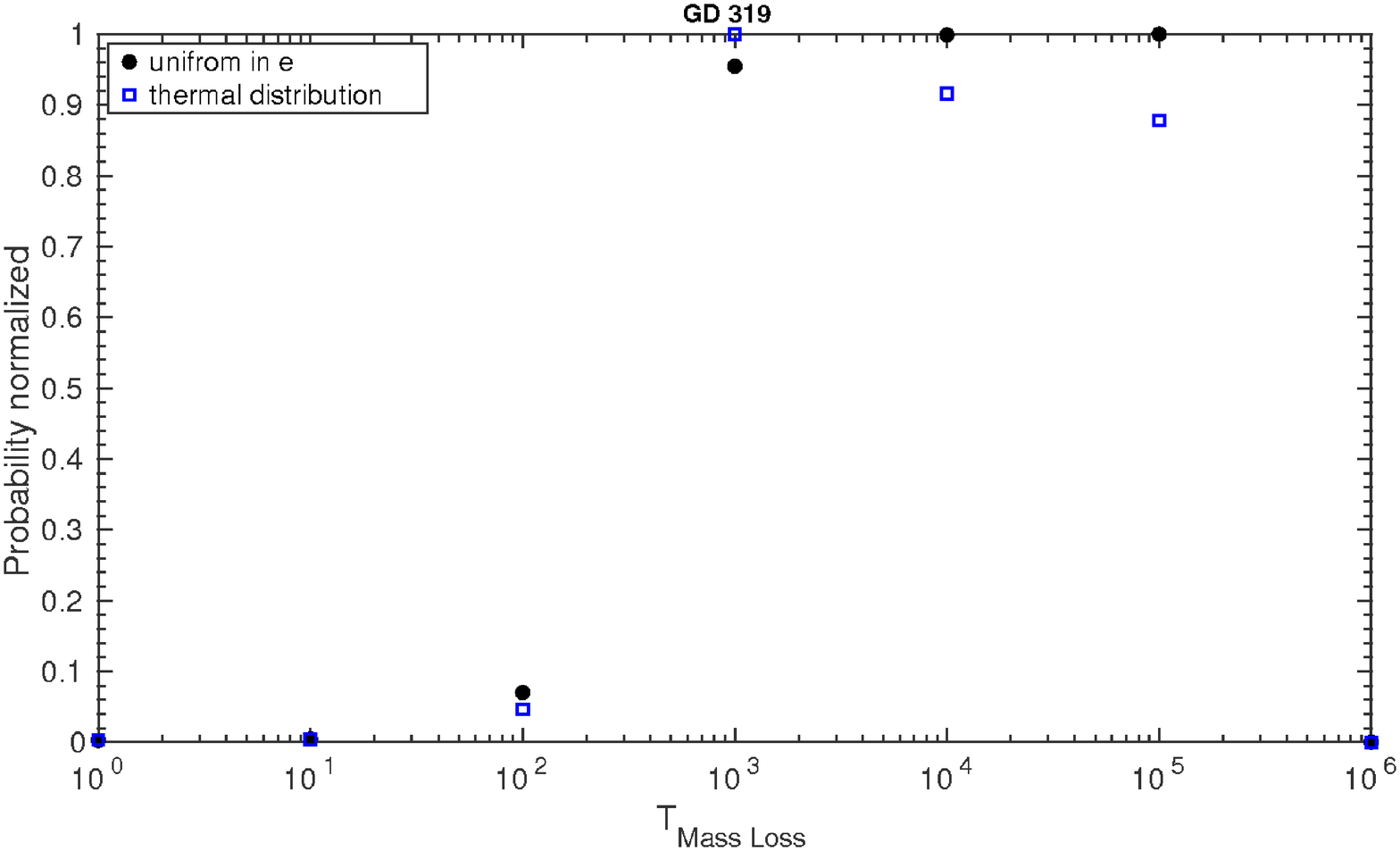}\caption{\label{fig:exponential-mass-loss_GD319}Upper plot: the probability
of the mass loss timescale from the observed GD 319 system for the
exponential mass loss case. The black circles (blue squares) correspond
to uniform distribution (thermal distribution) of initial eccentricity.
The most probable mass loss timescale for the uniform initial eccentricity
is $T_{{\rm ML}}=10^{5}{\rm yr}$, but almost identical value to $T_{{\rm ML}}=10^{4}{\rm yr}$.
For the thermal eccentricity distribution is $T_{{\rm ML}}=10^{3}{\rm yr}$.
Bottom plot: Is the normalized version of the upper plot to the maximal
value for each distribution. It is clear that mass loss timescales
of $10^{3}-10^{5}{\rm yr}$ are favorable compared to $1-10^{2}{\rm yr}$
and $10^{6}{\rm yr}.$ }

\end{figure}

In this subsection we present the relative likelihood of mass-loss
timescales for the exponential mass loss function for GD 319. We follow
the same procedure as described in subsection \ref{subsec:LinearMassLossGD319}.
Figure \ref{fig:exponential-mass-loss_GD319} presents the probability
(upper) and the normalized probability to the maximal value (bottom).
We find that the most probable mass loss timescale for the uniform
initial eccentricity distribution is $T_{{\rm ML}}=10^{5}{\rm yr}$.
It is more probable by an approximate factor of $500,250,14.3$ for
$1{\rm yr},10{\rm yr}$ and $100{\rm yr}$ mass loss timescales respectively.
While it is approximately identical to the probability of $T_{{\rm ML}}=10^{4}{\rm yr}$,
with a factor of $1.001.$ and similar probability to the mass loss
timescales of $10^{3}{\rm yr}$ with only a factor of $1.04$ separates
their likelihood. In our calculation $10^{6}{\rm yr}$ have zero probability.

For the thermal distribution of the initial eccentricity the most
probable mass loss timescale is $T_{{\rm ML}}=10^{3}{\rm yr}$. It
is more likely by approximate factors of $333.3,250$ and $21.7$
than the $1{\rm yr},10{\rm yr}$ and $100{\rm yr}$ mass-loss timescales
cases, respectively. While it is similar for the mass loss timescales
of $10^{4}{\rm yr}$ and $10^{5}{\rm yr}$. In this case the $10^{6}{\rm yr}$
timescale has zero probability.

\section{Discussion}

\label{sec:Discussion}

In order to calculate the results in sections \ref{sec:Wolf1130}
and \ref{sec:GD319} we made use of several assumptions, which we
discuss below.
\begin{itemize}
\item We assumed that the third companion with similar proper motion is
a bound companion. This is a commonly used assumption in the study
of wide binaries. If the companion was originally bound but the triple
was disrupted due to mass-loss, its separation would grown in time,
and would far exceed the observed separation after only a few $10^{4}-10^{5}$
yrs, much longer than the cooling age of the observed WD which is
$\sim3.4{\rm Gyr}$ for Wolf 1130 \citep{Mace2018}.
\item Mass loss timescale. We assumed the mass loss timescale from the inner
binary during the CEE is equal to the mass loss timescale from the
outer binary. This is not necessarily true when the mass ejected from
the envelope have a wide velocity spread. Our ignorance on the ejection
mechanism required us to make this simplifying assumption. Nevertheless,
if we assume the mass ejected is ejected with velocities similar to
escape velocities from a giant star, roughly $v\sim30{\rm kms^{-1}}$
and the spread of the ejected velocity is $\sim10\%$, namely $\Delta v\sim3{\rm kms^{-1}}$
we get the crossing time for the total mass of the ejected envelope
to cross the systems beyond the outer companion is 
\begin{equation}
\Delta T\approx a\times\frac{\Delta v}{v^{2}}=\left(\frac{a}{1000{\rm AU}}\right)\times15.8{\rm yr,}
\end{equation}
and therefore we can not constrain CE-ejection timescales shorter
than this range. However, the most likely timescales we find far exceed
this short timescale, and therefore does not affect our conclusions.
\item In this paper we consider two very different mass loss functions:
linear in time (\ref{eq:linear mass loss}) and exponential in time
(\ref{eq:exponential}). We find that regardless of the chosen functional
form of the mass-loss we find similar mass loss timescale constraints,
suggesting the overall method is generally robust to the mass loss
function.
\item We only analyzed the data of two specific systems. Given the small
statistics, in principal it is possible that in these two systems
a third companion survived a shorter timescale mass-loss stage than
the most probable one we find. Therefore although our results are
suggestive of long timescales for CE-ejection, more systems should
be analyzed in order to further support or refute these preliminary
results.
\item Throughout the paper we assumed the mass-loss is isotropic. An unisotropic
mass-loss, possibly expected in CE-ejection \citep{Taam2010} might
somewhat change the results, but is beyond the scope of this paper
focusing on the presentation of the basic method and its initial testing.
\end{itemize}
In this paper we make use of a third wide-orbit companion to constrain
the mass-loss timescale. We should note that a somewhat related idea
was used by \citet{El-Badry2018} to constrain natal kicks of WD,
although they used a generalized statistical analysis, and not accounted
for the detailed evolution of each system. We also note that they
did not consider the treatment of prompt mass-loss which is relevant
for their wider systems, and it is therefore possible that the cut-off
they find in wide-separation is not related to WD kicks, but in fact
corresponds to the range where the mass-loss timescale from AGB stars
becomes comparable or shorter than the orbital period of the wide
companions, as we discussed in this paper.

\section{Summary}

\label{sec:Summary}

Common-envelope evolution plays an important role in the evolution
of interacting binaries, but it is not well understood. Various different
models were suggested to explain the envelope-ejection process, some
of which differ in the the expected timescale involved in the ejection.
Here we proposed a method to constrain the CE mass-loss timescales
in CE-binaries, making use of CE-binaries with wide-orbit third companions.
The orbits and survival of such third companions strongly depends
on the timescale of the mass-loss and can therefore be used to con
train them. We describe the orbital evolution due to mass-loss, and
apply our method on two test cases of observed post-CE binaries which
are part of wide triple systems, Wolf 1130 and GD 319. We find the
most probable mass-loss timescales for these systems of $10^{3}-10^{5}{\rm yr}$
are longer than the timescales expected in most of the suggested models
for CE-ejection, but they could be consistent with CE-ejection through
dust-driven winds \citet{Glanz2018}. These results are currently
based only on these two test-cases, and depend to some extent on the
specific assumptions taken, and further analysis of additional systems
is desired on order to provide further support, or refute them. Nevertheless,
the general method suggested is robust, and can be used in the future
on large samples. 

Though based on two cases, the long CE-ejection timescales we infer
do suggest significant challenge to most of the CE-models suggest
to-date and are in tension with their expectations.

\section*{Acknowledgements}

We acknowledge support from the ISF-ICORE grant 1829/12. The authors
acknowledge the University of Maryland supercomputing resources (http://hpcc.umd.edu)
made available for conducting the research reported in this paper.





\bsp	
\label{lastpage}
\end{document}